\documentclass{article}

\PassOptionsToPackage{numbers, compress}{natbib}
\usepackage[final]{neurips_2022_ml4ps}

\usepackage[utf8]{inputenc} 
\usepackage[T1]{fontenc}    
\usepackage{hyperref}       
\usepackage{url}            
\usepackage{booktabs}       
\usepackage{amsfonts}       
\usepackage{nicefrac}       
\usepackage{microtype}      
\usepackage{xcolor}         

\usepackage{amsmath}
\usepackage{microtype}      
\usepackage{color}
\usepackage{graphicx}
\usepackage{graphbox}
\usepackage{wrapfig}

\title{Towards solving model bias in cosmic shear\\ forward modeling}

\author{%
  Benjamin Remy\thanks{Corresponding author: benjamin.remy@cea.fr},~~ Fran\c{c}ois Lanusse~~and~~Jean-Luc Starck\\
  Université Paris-Saclay, Université Paris Cité, CEA, CNRS, AIM, 91191, Gif-sur-Yvette, France
    \\
}

\begin{document}

\maketitle

\begin{abstract}
As the volume and quality of modern galaxy surveys increase, so does the difficulty of measuring the cosmological signal imprinted in galaxy shapes. Weak gravitational lensing sourced by the most massive structures in the Universe generates a slight shearing of galaxy morphologies called cosmic shear, key probe for cosmological models. Modern techniques of shear estimation based on statistics of ellipticity measurements suffer from the fact that the ellipticity is not a well-defined quantity for arbitrary galaxy light profiles, biasing the shear estimation. We show that a hybrid physical and deep learning Hierarchical Bayesian Model, where a generative model captures the galaxy morphology, enables us to recover an unbiased estimate of the shear on realistic galaxies, thus solving the model bias.

\end{abstract}

\section{Introduction}

Over the next couple of years, a new generation of wide-field optical galaxy surveys will be coming online, starting with the ESA Euclid mission \citep{Laureijs2011} and the Vera C. Rubin Observatory Legacy Survey of Space and Time \citep{Ivezic2019}, soon followed by the Roman Space Telescope \citep{Spergel2015}. At the core of the cosmology science that these surveys will enable is the measurement of the weak gravitational lensing effect, a very subtle shearing of the apparent images of distant galaxies in the presence of massive structures along the line of sight. On cosmological scales, this effect, also known as cosmic shear, is an invaluable probe of the total matter content of the Universe, and will play a key role in further constraining and testing our cosmological model over the next decade.

Despite over 20 years of research in this field, unbiased measurements of gravitational shear from observed galaxy images remains one of the most challenging aspects of the scientific exploitation of these surveys. The prevailing strategy for measuring shear from data has been to build estimators of the ellipticity of galaxy light profiles, due to the fact that one can in theory relate how a galaxy ellipticity $e$ changes under a gravitational shear $\gamma$, as $e_\text{obs} \simeq e_{i} + \gamma$. Under the assumption that galaxies should have no preferential intrinsic ellipticity, a simple estimator for $\gamma$ can therefore be constructed from $\hat{\gamma} = \langle e_\text{obs} \rangle = \langle e_i \rangle + \gamma$. However, to this day, no shear measurement algorithm has proven to be unbiased. 

In fact, state-of-the-art approaches for shear measurement \cite{huff2017metacalibration, sheldon2017practical} no longer attempt to build an unbiased shear estimator from images, but rather calibrate the shape measurement methods estimating a multiplicative and an additive bias. There are many factors contributing to these biases, but a fundamental issue comes from the fact that the ellipticity $e$ is not a well defined quantity for arbitrary galaxy light profiles. 

In this paper we propose a new paradigm for cosmic shear estimation where we no longer rely on ellipticities anymore. We propose to forward model the galaxy images at the pixel level, similarly to \cite{schneider2015hierarchical}, but using non-parametric galaxy light profiles. The main idea is to train a deep generative model to learn galaxy morophologies, which can then be used as part of a hierarchical Bayesian model, a hybrid deep learning and physical model, simulating the observations and used to infer the shear.

The purpose of this paper is first to demonstrate that using parametric light profiles as part of the forward model induces a bias in the shear estimation, which we call a \textit{model bias}, and then to show that this bias can be removed by using a deep generative model for the morphology. We show in this preliminary study, that using a hybrid deep learning-physical Bayesian forward model can not only solve the model bias, but also provide uncertainty quantification to the shear estimation.

\begin{figure}[h]
    \centering
    \begin{tabular}{cc}
        
        \includegraphics[width=0.47\textwidth]{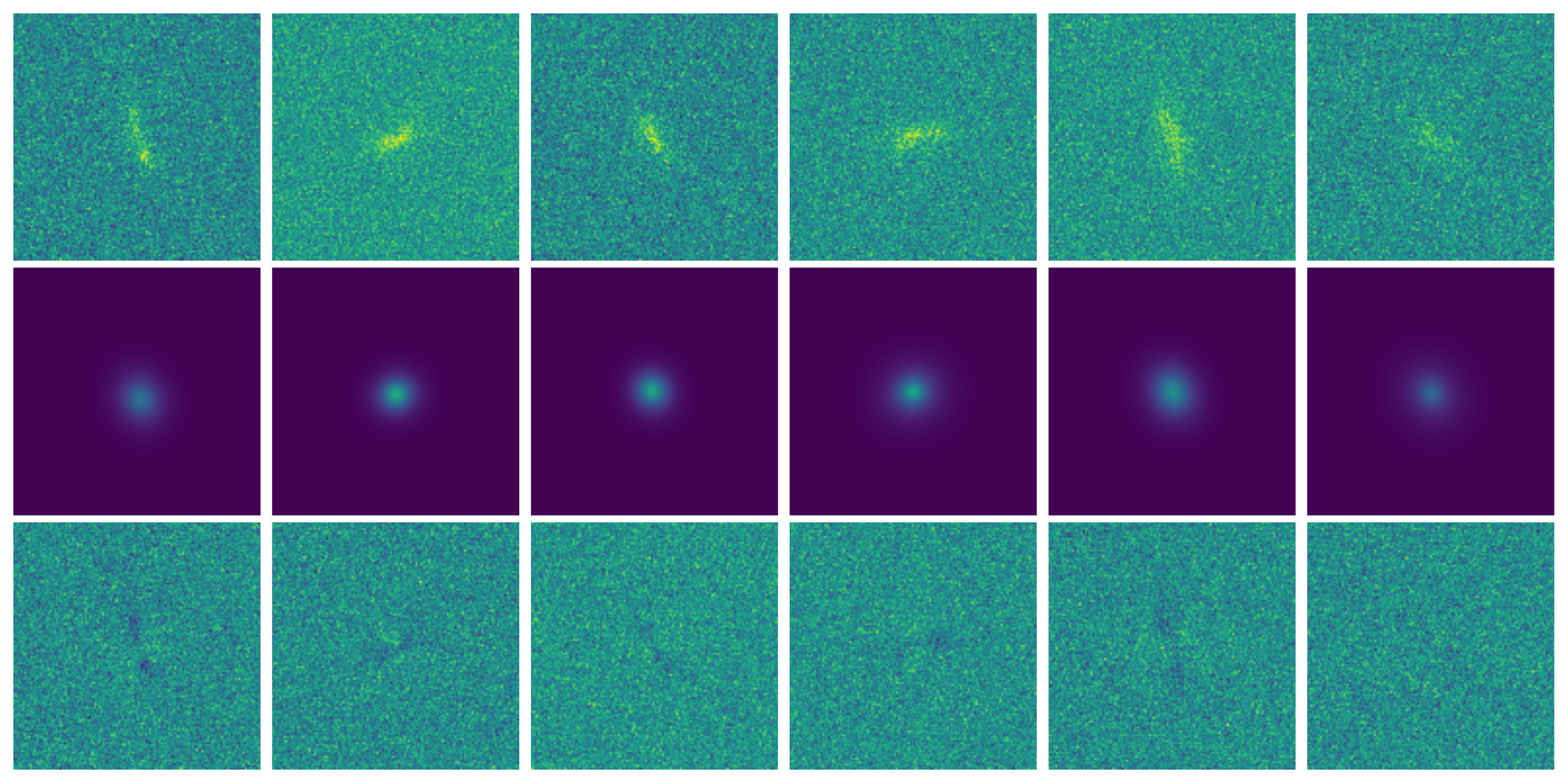} & \includegraphics[width=0.47\textwidth]{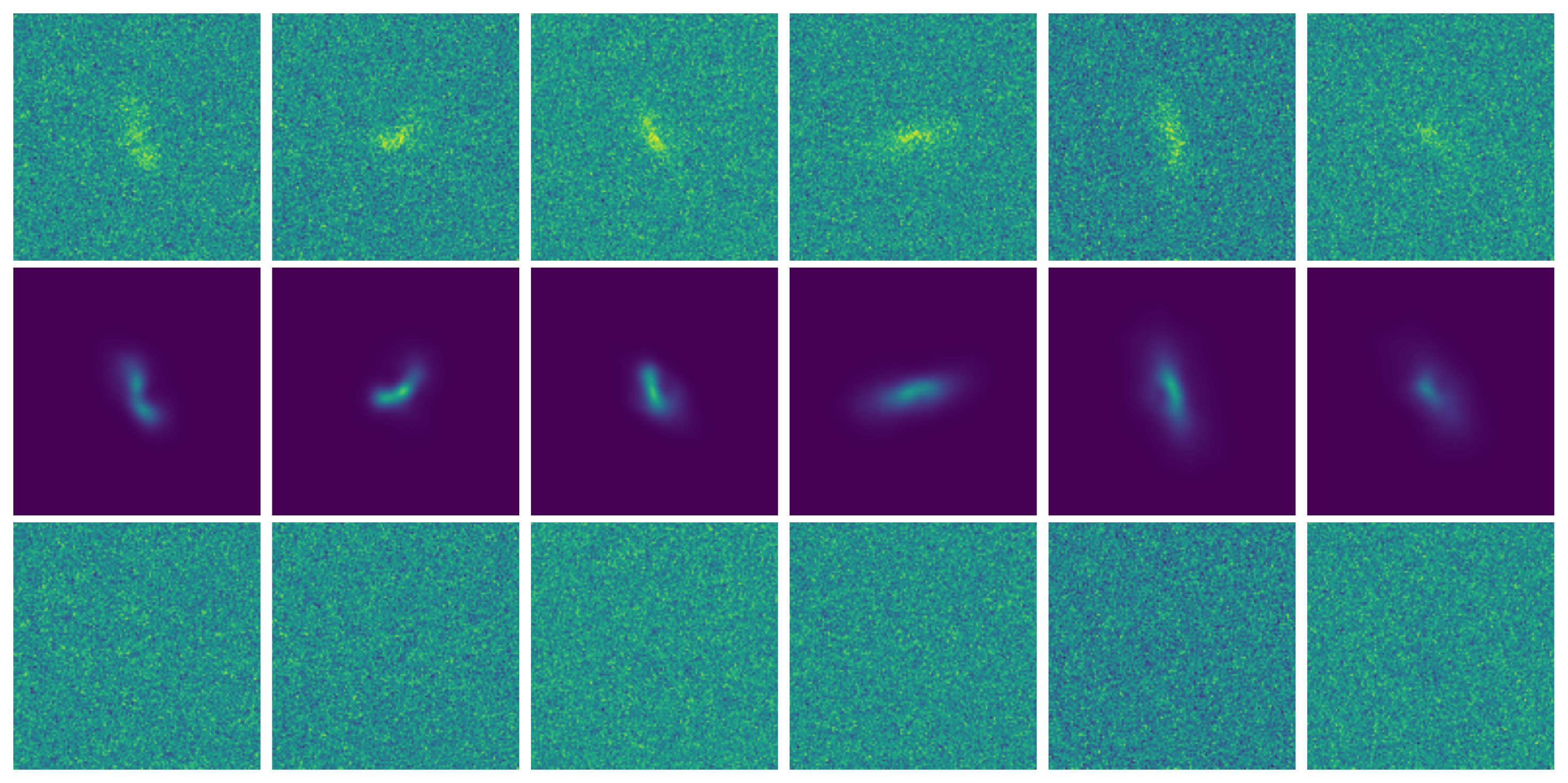} \\  

        (a) \small{Realistic galaxies fitted by a parametric profile} & (b) \small{Realistic galaxies fitted by a deep morphology model}
    \end{tabular}
    \caption{The first row of images correspond to observed galaxies. The second row corresponds to the light profiles Maximum a Posteriori fits, convolved with their PSFs. The third row shows  residual images between the observations and the model fits. Column (a) is using Sérsic model as light profile and column (b) is using \texttt{deepgal} model as light profile.
    }
    \label{fig:MAP}
\end{figure}

\section{Forward modeling weak lensing observations at the pixel level} \label{sec:models}

From a Bayesian perspective, making the joint inference of cosmic shear and galaxy morphologies means estimating the \textit{joint posterior} distribution $p(\mathcal{G}, \gamma | \mathcal{D})$, where $\mathcal{G} = \{z_i\}_{i=1\dots N}$ is the set of morphology parameters and $\mathcal{D} = \{d_i\}_{i=1\dots N}$ is the set of observed data. The Bayes identity tells us:
\begin{equation}\label{eq:bayes}
    p(\mathcal{G}, \gamma | \mathcal{D}) \propto p(\mathcal{D} | \mathcal{G}, \gamma) p(\mathcal{G}, \gamma),
\end{equation}
where $p(\mathcal{D} | \mathcal{G}, \gamma)$ is the data \textit{likelihood} given by the forward model described bellow. Since the intrinsic morphology of galaxies are independent of the cosmic shear, then $p(\mathcal{G}, \gamma) = p(\mathcal{G})p(\gamma)$, where $p(\mathcal{G})$ and  $p(\gamma)$ are respectively the \textit{prior} distributions of the morphological parameters and of the cosmic shear.

\begin{wrapfigure}[12]{r}{0.4\textwidth} 
    \vspace{-25pt}
    \centering
    \includegraphics[width=0.35\textwidth]{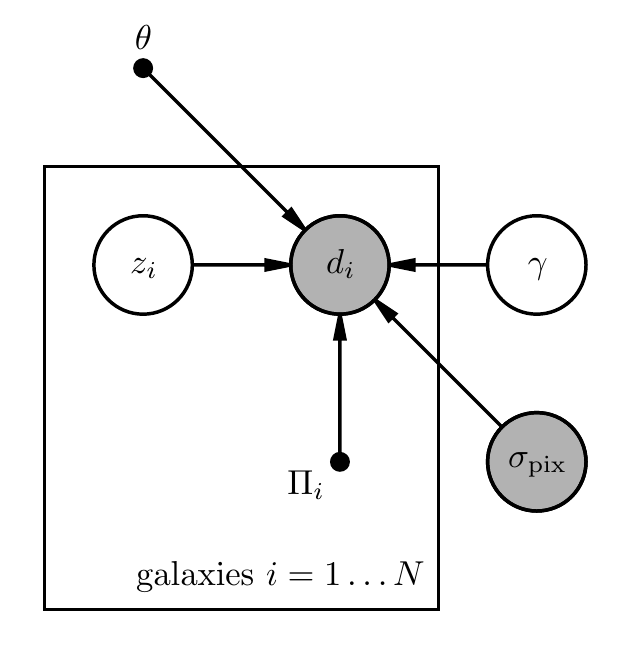}
    \vspace{-10pt}
    \caption{Probabilistic Graphical Model.
    }
    \label{fig:pgm}
\end{wrapfigure}

We show the probabilistic graphical model in \autoref{fig:pgm}. Nodes denote random variables and arrows conditional relation between them. White nodes corresponds to random variable on which we do the inference on, conditioned on the shaded nodes corresponding to observed variables. Dot nodes correspond to fixed variables, e.g. neural network weights $\theta$. In this paper we compare two forward models: (1) using Sérsic law  and (2) using the \texttt{deepgal} model to generate the galaxy light profile. Once the galaxy light profiles are drawn, we shear images with a constant value $\gamma$ shared by all the galaxies. Each sheared galaxy $i$ is then convolved by its Point Spread Function (PSF) $\Pi_i$. Assuming Gaussian noise corruption, the resulting image is therefore evaluated by a Gaussian likelhood of variance $\sigma_\text{pix}^2$.

\textbf{Sérsic model~~}
The light profile of the Sérsic model is given by the following parametrization: 
$I(r) = F \times I_0 \exp ( -b_n [( \frac{r}{r_\text{hlr}})^{\frac{1}{n}} -1] ),$
where $r_\text{hlr}$ is the half-light-radius, $I_0$ is a normalization factor such that the integration of $I$ over all the space is $F$, which corresponds to the flux of the galaxy. $n$ is called the Sérsic index and $b_n$ is a dimensionless constant depending on $n$. The galaxy is then sheared with parameters $e=(e_1, e_2)$ corresponding to its ellipticity. Finally, since the real galaxy may not be perfectly centered in the postage stamp, we add two shift parameters $s_x$ and $s_y$. Therefore, each galaxy $i$ has a set of parameter $z_i = \{n, r_\text{hlr}, F, e_1, e_2, s_x, s_y\}_i$. Prior distributions and parameters follow empirical distributions of the COSMOS shape catalog and are described in \autoref{sec:cosmos-prior}.

\textbf{\texttt{Deepgal} model~~} \texttt{deepgal} is a VAE-based model, proposed in \citep{lanusse2021deep}, generating complex galaxy light profiles. In this model, the observations are generated following a random process mapping a latent space representation $z_i \in \mathbb{R}^{16}$ to the observed image $d_i$ following the probabilistic model illustrated in \autoref{fig:pgm}. The neural network weights $\theta$ of the VAE are trained maximizing the Evidence Lower Bound (ELBO) of the data. Moreover, another generative model was trained on top of the VAE to learn latent space distribution using Masked Auto Regressive Flows \cite{papamakarios2017masked}. This model was trained on the HST/ACS COSMOS survey postage stamps.

In the weak lensing regime, the shear is very small, therefore we can assume that it does not change the morphological statistics of the training sample. Thus, the model can be trained once, a priori, over a large catalog of galaxies and be used to perform the joint inference of the latent variables and the cosmic shear. For both light profile models, we denote the generated image $\mathcal{I}_0$.

\textbf{Cosmic shear~~} In this model every galaxy image is sheared by a constant value $\gamma = \gamma_1 + i \gamma_2$ shared by all the galaxies. The image shearing transform, denoted $\mathcal{I}_0 \otimes \gamma$, is given by the following pixel coordinate change: $\big(\begin{smallmatrix}
  x^\prime \\
  y^\prime 
\end{smallmatrix}\big) = S(\gamma) \big(\begin{smallmatrix}
  x \\
  y 
\end{smallmatrix}\big)$,
such that $S(\gamma) = \frac{1}{\sqrt{1-|\gamma|^2}}
                    \big(\begin{smallmatrix}
  1+\gamma_1 & \gamma_2 \\
  \gamma_2 & 1-\gamma_1
\end{smallmatrix}\big)$.

\textbf{Convolution with the PSF~~}
The Point Spread Function (PSF) model is assumed to be known and exact for each galaxy. We draw the PSF evaluated at the galaxy $i$ position, denoted by $\Pi_i$, in Fourier space and then multiply it to the Fourier image of the light profile before returning in the real space, resulting in an image $\mathcal{I}$, such that $\mathcal{\tilde I} =  (\mathcal{\tilde I}_0 \otimes \gamma) \cdot \tilde \Pi$.

\textbf{Noise corruption~~}We assume a known fixed gaussian noise identical for all the galaxies. For simplicity, all of the galaxies in our observations are corrupted by a constant Gaussian noise of variance $\sigma_\text{pix}^2$, as well as for our forward model. In a more realistic scenario where we would have information about the noise correlation properties for each postage stamp, one could consider a noise model per galaxy. The likelihood is therefore given by 

\begin{equation}
    p(\mathcal{D} | \mathcal{G}, \gamma) \propto \prod_{i=1,\dots, N} \mathcal{N}\left( d_i \Big|  \mathcal{F}^\dagger\left[ \left(\mathcal{\tilde I}_{0,i} \otimes \gamma \right) \cdot \tilde \Pi_i \right], \sigma^2\right), ~~~\mathcal{F}^\dagger~ \text{is the inverse Fourier operator}.
\end{equation}

\vspace{-10pt}

\section{Illustration on realistic data}

\begin{figure}[h]
    \centering

    \includegraphics[width=\textwidth]{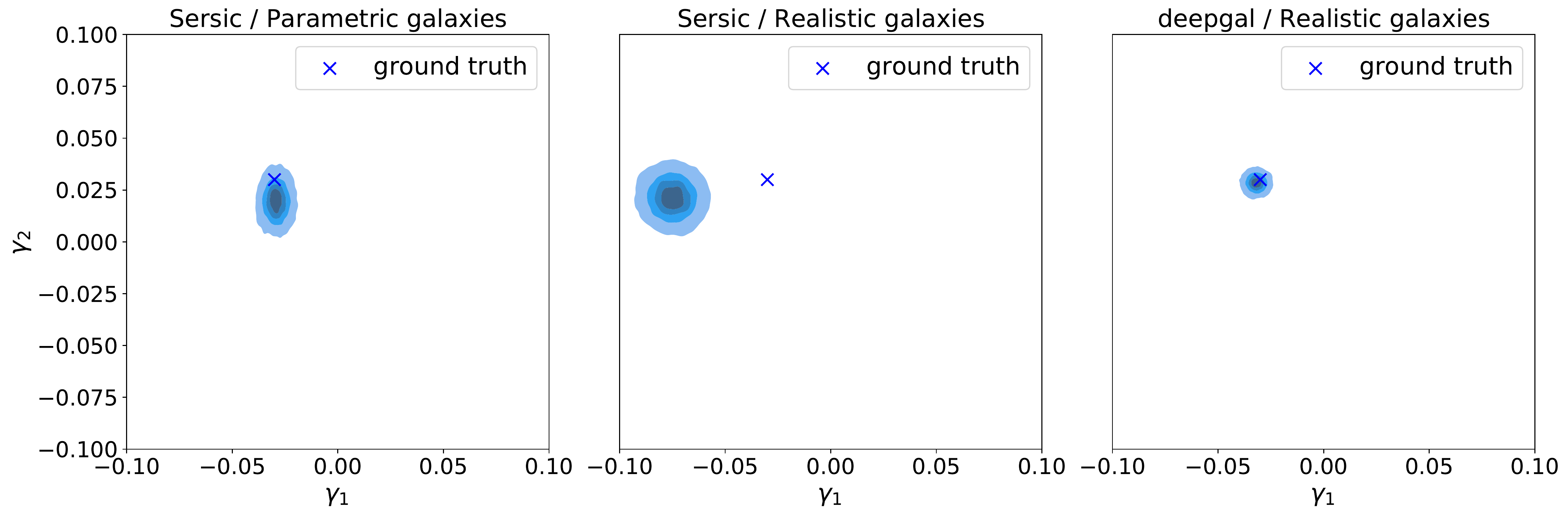}
    \caption{Shear marginal posterior contours against ground truth $\gamma=(-0.03, 0.03)$. The left panel shows the posterior using the Sérsic light profile over the parametric observations. The middle panel shows the posterior using the Sérsic light profile over the realistic galaxies. The right panel shows the posterior using the \texttt{deepgal} model over the realistic galaxies.
    }
    \label{fig:bias}
\end{figure}

\vspace{-10pt}

\subsection{Simulating observations} \label{sec:data}

We designed a toy model where galaxies are assumed to be already detected, deblended and centered in postage stamps, so that we only need to simulate the light profile, shearing transform and convolution with the PSF. In order to demonstrate the model bias, we distinguish two sets of observations. (1) generated with parametric galaxy light profiles using a Sérsic model and (2) using realistic galaxies based on the galaxy postage stamps from the COSMOS HST Advanced Camera for Surveys (ACS) field \citep{koekemoer2007cosmos, scoville2007cosmic_a, scoville2007cosmos_b}. We used the postage stamps from \cite{leauthaud2007weak} and Sérsic fit parameters compiled in \cite{mandelbaum_rachel_2012_3242143}. Because the unconvolved galaxies are very noisy, we used the Auto-Encoder from \citep{lanusse2021deep} to get denoised version of unconvolved galaxies, therefore this is not a demonstration on real galaxies yet. However, the purpose of this paper is to demonstrate the model bias, so we find that realistic galaxy morphologies that cannot be captured by analytic profiles are sufficient to illustrate the argument.

Each set of observation is a field of 256 galaxies, generated from the COSMOS catalog, with the following cuts: $n > 0.4$, $r_\text{hlr} > 0.2$ and $23.5 < i^+ < 25.2$ so rather faint galaxies. That way the entire galaxy is in the postage stamp, is not too peaky and not too resolved to be close to a Sérsic model. Every postage stamps were corrupted by a Gaussian noise of standard deviation $\sigma_\text{pix}=0.01$. In order to simulate shearing, we removed any shear present in the selected sample from the COSMOS catalog, by adopting the same strategy than in the GREAT3 Challenge \citep{mandelbaum2014third}, i.e. doubling each galaxy, and rotating the doubled galaxies by 90°, as well as their PSF, to cancel the overall sample shear and then apply the same constant shear to every galaxy $\gamma=(-0.03, 0.03)$.

\subsection{Approximate inference with mean-field VI}

In order to estimate the posterior distribution $p(\mathcal{G}, \gamma | \mathcal{D})$, we relied on approximate inference using mean-field Variational Inference (VI), where we approximate the posterior distribution using a parametric distribution $q_\phi(\mathcal{G}, \gamma) = q_{\phi_\gamma}(\gamma) \prod_i\prod_j q_{\phi_{i,j}}(z_{i,j})$,
parameterized by $\phi$. We used Normal distributions $q_{\phi_{i,j}}$ to approximate the posteriror of parameters $i$ of the $j^\text{th}$ galaxy, so the parameters $\phi_{i,j} = (\mu_{i,j}, \sigma_{i,j})$ correspond to means and standard deviations. Similarly, $q_{\phi_\gamma}$ is a 2d-diagonal multivariate Gaussian. The parameters $\phi_{i,j}$ and $\phi_\gamma$ are then optimized maximizing the ELBO:
\begin{equation}
    \phi^* = \arg \max_\phi ~~ \mathbb{E}_{(\mathcal{G}, \gamma) \sim q_\phi} \left[ \log p(\mathcal{G}, \gamma, \mathcal{D}) - \log q_\phi(\mathcal{G}, \gamma) \right]
\end{equation}
This remains a very simplistic approach, which does not assume any correlation between variables. A more accurate model would condition the morphological parameters on the shear. In principle, the most accurate method would be to run a Markov Chain Monte Carlo, but several experiments showed that sampling based-methods are too computationally expensive for such high-dimensional problem.
\subsection{Identifying and solving the model bias}

We ran three sets of experiments to demonstrate the model bias. First fitting the fully parametric model to the parametric observations. Then fitting the same parametric model to the realistic galaxies and then fitting the hybrid model to the realistic galaxies. We used a Sérsic law as light profile as proposed in \cite{schneider2015hierarchical}. One could argue that there exist more complex parametric profiles such as the bulge+disk model which could reduce the aforementioned bias, but we propose here a model which does not rely on ellipticities and in principle can capture any complexity of galaxy morphologies.

First, as illustrated in \autoref{fig:bias} (a), the inferred shear using the parametric model on the parametric observations is unbiased. This was indeed expected since the observations and the model are exactly the same. Secondly, \autoref{fig:bias} (b) shows that the parametric model induces a bias in the shear estimation when the galaxy morphologies are no longer elliptical. Besides, one can observe the non-zero residuals in the \autoref{fig:MAP} fit to real galaxies, illustrating the model discrepancy when applied to complex morphologies. Finally, \autoref{fig:bias} (c) shows that when we substitute the parametric light profile model by the \texttt{deepgal} model, the shear estimation becomes unbiased. One can notice that the uncertainties using the generative model are smaller than for the Sérsic model, but these contours are not representative of the error bars due to the mean-field approximation.

\section{Discussion and conclusion}

We have presented in this work the necessity of an accurate light profile model for forward modeling of the cosmic shear. We have shown that a hybrid physical and deep learning HBM has the potential to solve the model bias, being able to capture the complex morphology of galaxies and simulating physical processes such as shearing transform and convolution by the PSF. We identify many directions of research towards a real cosmological analysis by forward modeling. On the one hand, many assumptions could be relaxed by building a more complex hierarchical model, e.g. parametrizing the PSF model and jointly inferring it along with the shear. Concerning the detection and deblending of galaxies, similar work have already been proposed, e.g. BLISS \cite{hansen2022scalable}. On the other hand, accurate uncertainty quantification could be enabled by scaling MCMCs for HBMs or using neural density estimator to capture the shape of the joint posterior distribution.

\section{Acknowledgments}

BR acknowledges support by the Centre National d’Etudes
Spatiales (CNES). This work is supported by a public grant overseen by the French National Research Agency (ANR) through the program UDOPIA, project funded by the ANR-20-THIA-0013-01. This work was granted access to the HPC resources of IDRIS under the allocation 2022-102038 made by GENCI.

\section{Broader impact}

The methodology described here finds many useful applications within and outside of cosmology. Hierarchical Bayesian Models are very efficient tools for statistical inference but can suffer form model misspecification. Combining deep generative models and physical models seems to be a promising approach to solve this problem. We believe this work does not entail any negative consequences or ethical issues.


\bibliographystyle{unsrt}
\bibliography{references}

\appendix

\section{Appendix}

\subsection{Prior distributions and parameters for Sérsic light profile} \label{sec:cosmos-prior}
The following distributions were fitted from the COSMOS shape catalog \cite{leauthaud2007weak, mandelbaum_rachel_2012_3242143}.

\begin{itemize}
    \item $\log_{10}r_\text{hlr} \sim \mathcal{N}(-0.68, 0.30)$
    \item $\log_{10}n \sim \mathcal{N}(0.1, 0.39)$
    \item $\log_{10}F \sim \mathcal{N}(-1.97, 0.53)$
    \item $e_{1/2} \sim \mathcal{N}(0, 0.28)$
    \item $\gamma_{1/2} \sim \mathcal{N}(0, 0.09)$
    \item $s_{x/y} \sim \mathcal{N}(0, 1)$
\end{itemize}

\end{document}